\documentclass[12pt]{article}
\usepackage{pdproc,graphicx} 

\begin{document}

\renewcommand{\thefootnote}{\alph{footnote}}
  
\title{
 SUPERNOVA NEUTRINOS:\\[1mm]
 STRONG COUPLING EFFECTS OF WEAK INTERACTIONS}

\author{G.L.~FOGLI~$^{1,2}$, E.~LISI~$^{2*}$, A.~MARRONE~$^{1,2}$, 
A.~MIRIZZI~$^{2,3}$}
\smallskip
\address{ $^1$~Dipartimento di Fisica, Universit\`a di Bari,
Via Amendola n.176, 70126 Bari, Italy \\[+0.4mm]
$^2$~Sezione INFN di Bari, Via Orabona n.4, 70126 Bari, Italy\\[-0.3mm]
$^3$~Max-Planck-Institut f\"ur Physik, F\"ohringer Ring 6, 80805 M\"unchen,
Germany}

\centerline{\footnotesize $^*$Speaker. \tt eligio.lisi@ba.infn.it}
\abstract{In core-collapse supernovae, $\nu$ and $\overline\nu$ are initially subject to significant self-interactions 
induced by weak neutral currents, which may induce 
strong-coupling effects on the flavor evolution (collective transitions).
The interpretation of the effects is simplified when
self-induced collective transitions are decoupled 
from ordinary matter oscillations, as for the matter density profile
that we discuss.
In this case, approximate analytical tools 
can be used (pendulum analogy, swap of energy spectra).
For inverted $\nu$ mass hierarchy, the sequence of effects involves:
synchronization, bipolar oscillations, and  spectral split. 
Our simulations shows that the main features of these regimes are 
not altered when passing from simplified (angle-averaged) treatments to full, multi-angle
numerical experiments.}

\normalsize\baselineskip=15pt

\begin{figure}[b]
\center
\includegraphics[height=1.9in]{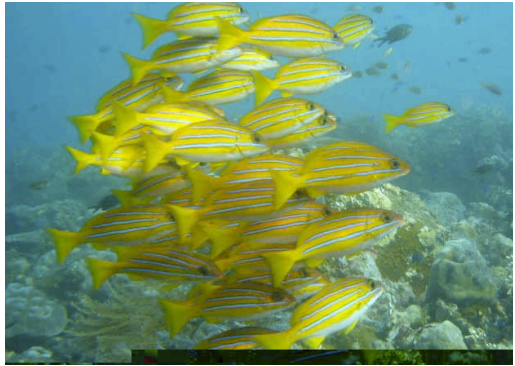}
\caption{In a school of fish, individuals often show a collective behavior. Very dense neutrino gases, like those emerging from a core-collapse supernova, 
might show analogous features in flavor space.} 
\label{school}
\end{figure}

\section{Prologue}

Densely packed inviduals often show a surprising, collective behavior (Fig.~\ref{school}). Recent developments suggest  
that neutrinos make no exception, despite the weakness of their  
self-interactions. Effects of $\nu$-$\nu$ forward scattering 
(via neutral currents) may be as important as the known effects
of $\nu_e$-$e^-$ forward scattering in matter (via charged currents), 
provided that the $\nu$ number density is very
high. The dense core of exploding Supernovae might
provide a possible environment where
the flavor evolution of $\nu$'s and $\overline\nu$'s can
show, indeed, highly nonlinear and strongly coupled effects.
\newpage

\begin{figure}[t]
\phantom{.}\vspace*{2mm}
\center
\includegraphics[height=2.5in]{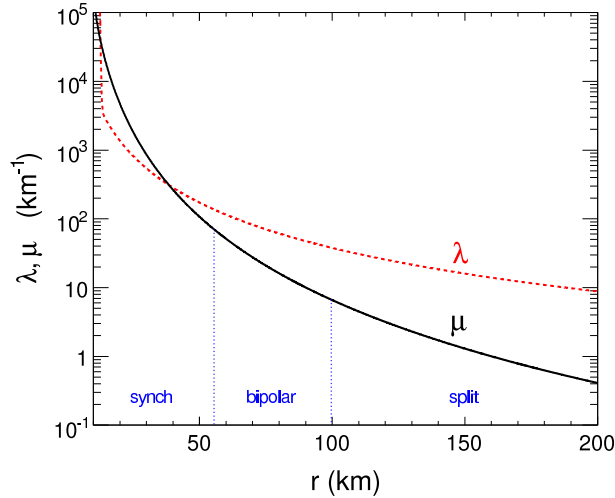}
\caption{Radial profiles of the neutrino self-interaction parameter
$\mu(r)=\sqrt{2}\,G_F\,(N+\overline N)$ and of the matter-interaction
parameter $\lambda(r)=\sqrt{2}\,G_F\,N_{e^-}$ adopted in this work,
in the range $r\in[10,\,200]$~km.}
\end{figure}
\vspace*{-3mm}

\section{Reference supernova model}

 Supernova $\nu$ oscillations are a very important tool to study astrophysical
processes and to better understand $\nu$ properties~\cite{Raffelt:2007nv}. 
After leaving the neutrinosphere, $\nu$ and $\overline\nu$ 
 undergo flavor oscillations triggered
by vacuum mass-mixing parameters and by ordinary (MSW) matter effects. Besides, 
in the first few hundred kilometers
neutrino-neutrino interactions may induce additional
important effects (depending on the neutrino mass hierarchy).
Self-interaction effects are expected to be non negligible when
$\mu(r) \sim \omega$, where
$\mu =  \sqrt{2} G_F (N_\nu(r) + \overline{N}_\nu(r))$ is the potential associated to the $\nu+\overline\nu$ background  
[analogous to the MSW potential $\lambda = \sqrt{2} G_F N_{e^{-}}(r)$], while
 $\omega=\Delta m^2/E$ is the largest vacuum oscillation frequency.
We neglect the smallest mass squares difference
$\delta m^2=m^2_2-m^2_1\ll \Delta m^2=|m^2_3-m^2_{1,2}|$, and consider a 
$2\nu$ mixing 
scenario governed by $\Delta m^2$ and the mixing angle $\theta_{13}$
($\Delta m^2 = 10^{-3}$~eV$^2$ and $\sin^2\theta_{13} = 10^{-4}$ for reference).
In the supernova context, $\nu_\mu$ and $\nu_\tau$ (shortly, $\nu_x$) behave similarly, and
we can generically consider two-neutrino $\nu_e \leftrightarrow \nu_{x}$ oscillations as
a reasonable approximation.

Figure 1 shows the radial profiles of the matter potential $\lambda(r)$ and of
the neutrino potential $\mu(r)$, and the approximate ranges where 
different collective effects occur: synchronization,
bipolar oscillation and spectral split. 
The nonlinearity of the self interactions induce 
collective transitions for small $r$, well before  the ordinary
MSW resonance, allowing a clear interpretation of the numerical simulations.
However, for matter profiles different from ours
(shallow electron density profiles~\cite{Duan:2006an}),
the MSW effects can be already operative 
around $O(100)$~km, in which case (not typical, and not considered here) they are entangled to the collective ones in a complicated way.
\newpage

We adopt normalized thermal spectra with $\langle E_e \rangle = 10$~MeV, 
$\langle \overline E_e \rangle = 15$~MeV, 
and $\langle E_x \rangle = \langle \overline E_x \rangle = 24$~MeV
for $\nu_e$, $\overline\nu_e$, $\nu_x$ and $\overline \nu_x$,
respectively. 
The emission geometry is based on the so called
 ``bulb model''~\cite{Duan:2006an} with spherical symmetry: 
neutrinos are assumed to be half-isotropically emitted from the neutrinosphere. 
Along any radial trajectory there is, therefore, a cylindrical symmetry.
As cylindrical variables one can choose
the distance form the supernova center $r$, and the angle $\vartheta$ between two interacting neutrino trajectories.
If the dependence on $\vartheta$ is
integrated out, one speaks of ``single-angle''  approximation, while the general situation
of variable $\vartheta$ is dubbed ``multi-angle'' case. The numerical simulation in the multi-angle case
is extremely challenging, since it typically
requires the solution of a large system ($\sim\!\!10^{M}$, $M\geq 5$) of coupled non-linear equations, after discretization of
the cylindrical coordinates.

\section{Equations of motion, pendulum analogy, and spectral split}

The propagation of neutrinos of given energy $E$ is studied through the Liouville
equation for the $2\times2$ neutrino density matrix in flavor basis. 
By expanding it on the Pauli and the identity matrices, the equations of motion
can be expressed in terms of two flavor polarization vectors, ${\bf P}(E)$ for
any neutrino and $\overline{\bf P}(E)$ for any antineutrino.
By introducing a vector ${\bf B}$ that depends on the mixing angle 
$\theta_{13}$, and a vector ${\bf D }= {\bf J - \overline{J}}$ that is the difference between the integral
over the energy of ${\bf P}$ and ${\bf \overline P}$, the equations of motion can be written as,
\begin{eqnarray}
\dot \mathbf{P}&=&\left(+\omega {\bf B}+\lambda {\bf z}+\mu {\bf D}\right)\times {\bf P}\ ,\label{Bloch1}\\
\dot \mathbf{\overline P}&=&\left(-\omega {\bf B}+\lambda {\bf z}+\mu {\bf D}\right)\times {\bf\overline P}\ ,\label{eqm}
\end{eqnarray}
see \cite{ourCollective} and references therein.
In the general case, the polarization vectors depend also on the
neutrino emission angle $\theta_{0}$ at the neutrinosphere
(the neutrino intersection angle $\vartheta$ can be expressed in terms of $r$ 
and of  $\theta_{0}$).
The $\nu_e$ survival probability
$P_{ee}$ is a function of the flavor polarization $z$-component,
$P_{ee}=1/2(1+P^z_f/P_z^i)$, where the $i$ and $f$ refer to the initial and final
state, respectively (analogously for $\overline\nu_e$).

Collective effects show up by aligning such polarization vectors
(in flavor space) close to each other. In the alignment approximation,  
the equations of motion for  ${\bf P}(E)$ and $\bar{\bf P}(E)$ can be reduced
to collective ones describing a classical, gyroscopic pendulum:
a spherical pendulum of unit length in a constant gravity field, characterized by a point-like massive bob spinning around the pendulum axis with constant angular momentum \cite{Hannestad:2006nj,Duan:2007mv}.
The pendulum inertia is inversely proportional to $\mu(r)$, while its
angular momentum depends on the difference of the integrated polarization
vectors $\bf J$ and $\bf \bar J$, see \cite{ourCollective}. 
The motion of a spherical pendulum is, in general, a combination of a precession
and a nutation. 

In the case of normal neutrino mass hierarchy,
the pendulum starts close to the stable, downward 
position and stays close to it, as $\mu$ slowly decreases 
collective effect gradually vanish.
In the inverted hierarchy case, the pendulum starts close to the ``unstable,''
upward position, being slightly tilted by an angle of $O(\theta_{13})$.
At small $r$, when $\mu$ is large (small pendulum inertia), the 
bob spin dominates  and the pendulum remains precessing in the
upward position as a ``sleeping top'' \cite{Duan:2007mv}, a
situation named synchronization~\cite{Pastor:2001iu,Hannestad:2006nj}.
As $\mu$ decreases with $r$, the pendulum inertia increases
and, unavoidably, for 
any $\theta_{13}\neq 0$, the pendulum fall occurs with subsequent  
nutations, the so called bipolar oscillations.
The increase of the pendulum inertia with $r$ reduces the amplitude of the nutations, and
bipolar oscillations are expected to vanish when self-interaction 
and vacuum effects are of the same size. 

While the bipolar regime comes to an end, 
self-interaction effects do not completely vanish, and a spectral split 
builds up:  a ``stepwise swap'' between the $\nu_e$ and $\nu_x$
energy spectra. The neutrino swapping can be explained by the conservation of the pendulum ``energy''
and of the lepton number~\cite{RaffSmirn}.
The lepton number conservation is related to  the constancy of $D_z=J_z-\overline J_z$,
that is a direct consequence of the equation of motion.
For a more detailed description of the pendulum analogy in our reference model,
the reader is referred to our work~\cite{ourCollective} and references therein.

\begin{figure}[t]
\phantom{.}
\vspace*{15mm}
\begin{minipage}{17.5pc}
\includegraphics[height=2.65in]{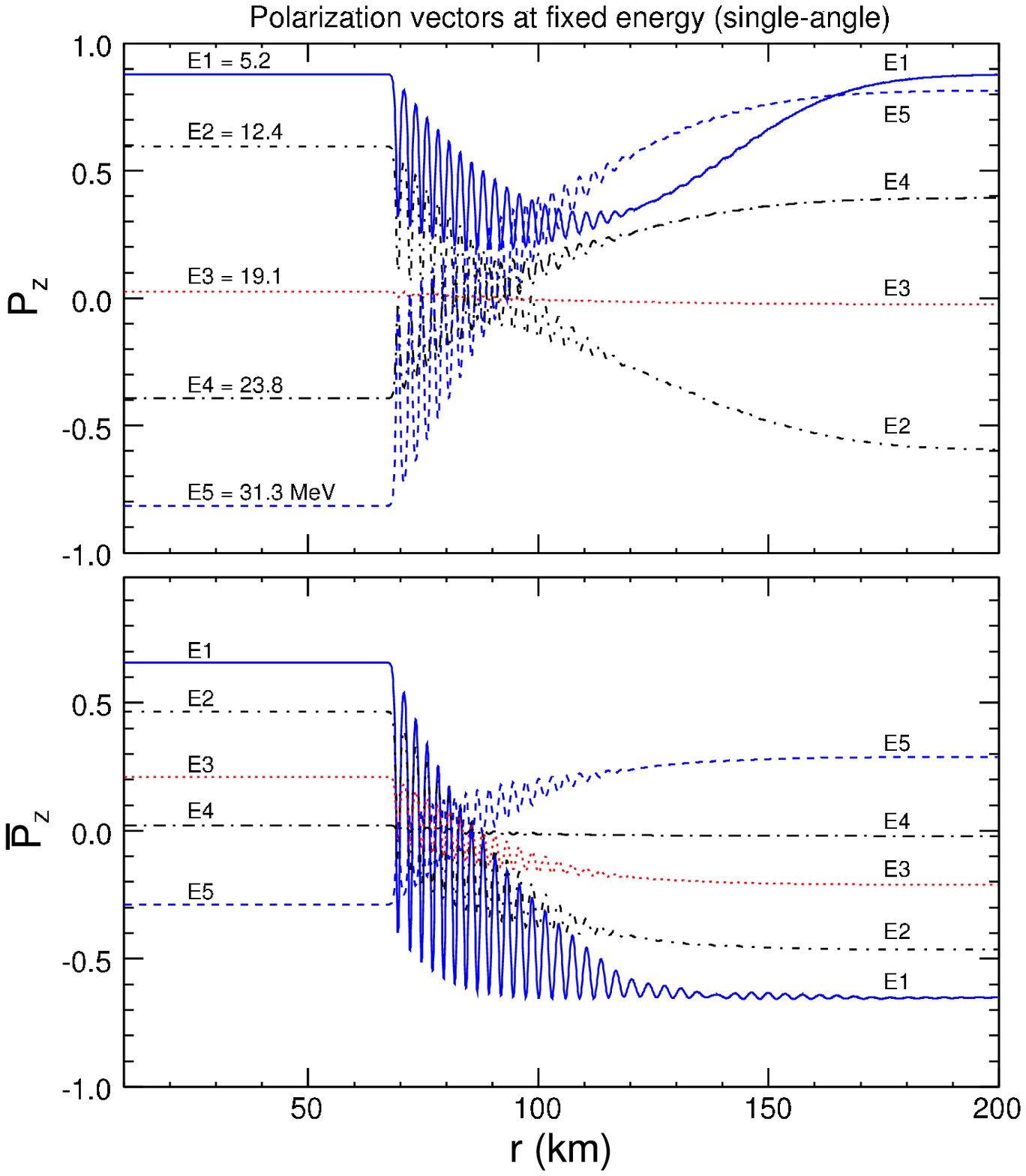}
\caption{Single-angle simulation in inverted hierarchy: $P_z$ (neutrinos)  and $\overline P_z$ (antineutrinos)
as a function of radius, for five energy values.
\label{fig4}}\end{minipage}
\hspace{1pc}%
\begin{minipage}{17.5pc}
\includegraphics[height=2.65in]{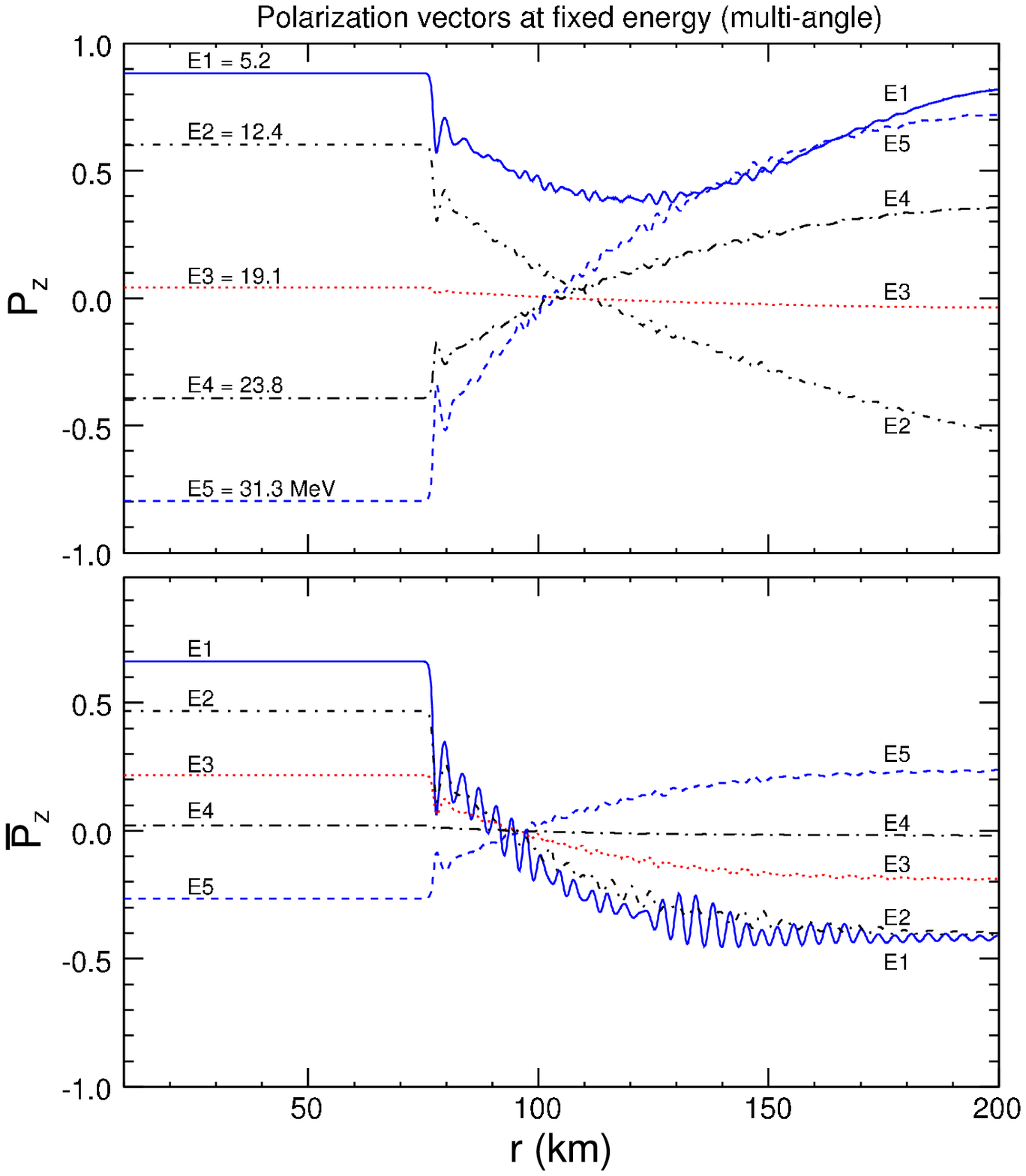}
\caption{Multi-angle simulation in inverted hierarchy: $P_z$ (neutrinos)  and $\overline P_z$ (antineutrinos)
as a function of radius, for five energy values.
\label{fig7}}
\end{minipage} 
\end{figure}

\section{Single- and multi-angle simulations: stability of results}

Figures 3 and 4 show the $z$-component of ${\bf P}$ and $\overline{\bf P}$, as a function of $r$ at different $E$ values, for single- and multi-angle simulations, respectively.
Bipolar oscillations start 
after a synchronization plateau, with equal periods for both $\nu$
and $\overline\nu$ at any energy, confirming the appearance  of collective features. Indeed,
the behavior of each $P_{z}$ and $\overline{P}_{z}$ depends 
essentially on its energy.
For neutrinos, Figure 3, the spectral split (inversion of
$P_z$) starts at a critical energy
$E_c\simeq 7$ MeV: the curve relative to $E<E_c$ ends up at the same initial value ($P_{ee} =1$),
while the curves for $E>E_c$ show the $P_z$
inversion ($P_{ee} = 0$). 
Neutrinos with an energy of $\sim 19$ MeV do not oscillate much, because this is roughly the energy for which
the initial $\nu_{e}$ and $\nu_{x}$ fluxes are equal in our scenario.
For $\overline\nu$, all curves show almost complete polarization reversal, except at very small energies (of few MeV, not shown).
Multi-angle simulations (Fig.~4) are similar, although with 
somewhat damped bipolar oscillations.

\begin{figure}[t]
\begin{minipage}{17.5pc}
\includegraphics[height=2.25in]{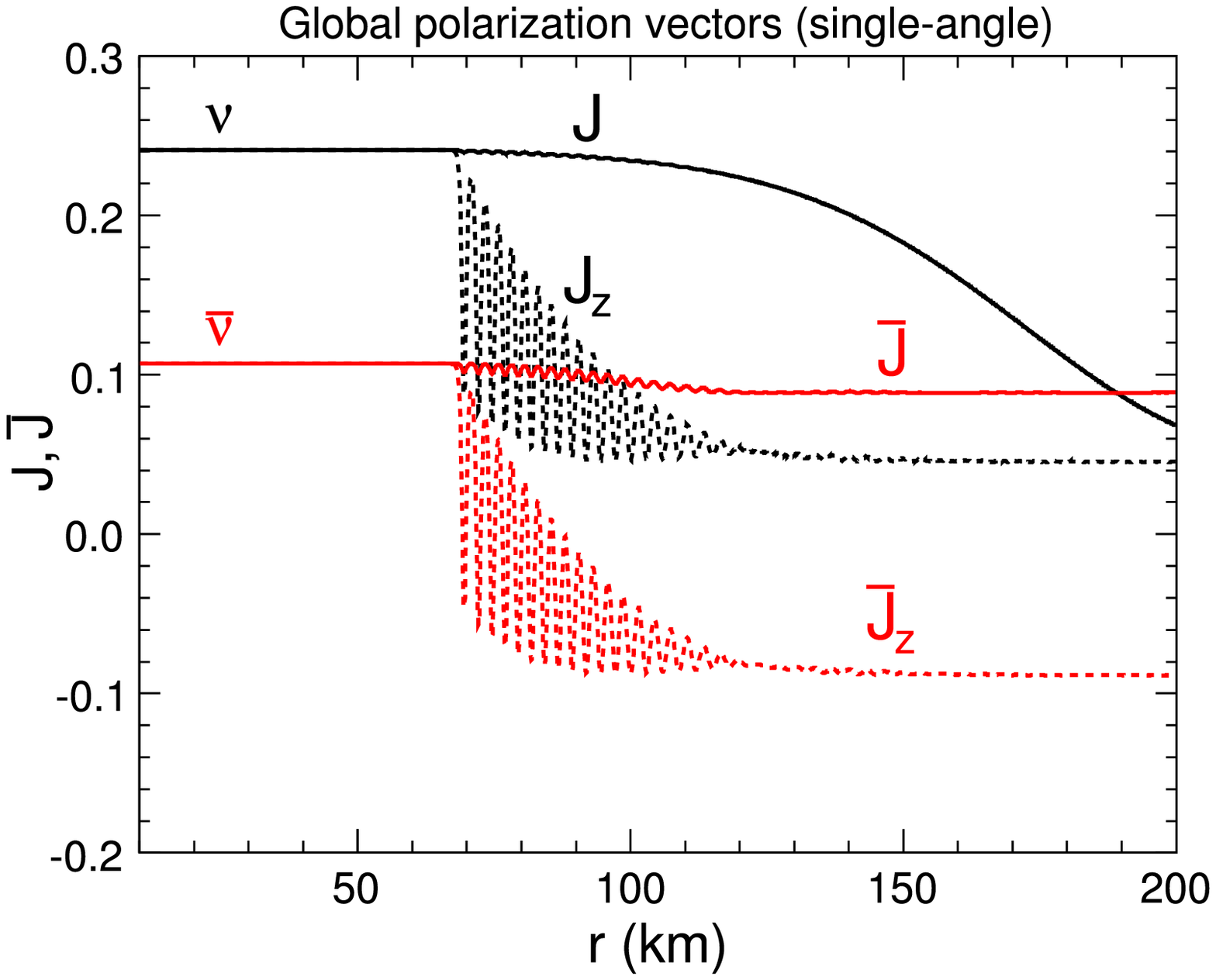}
\caption{Single-angle simulation in inverted hierarchy: modulus and $z$-component of
$\bf J$ and $\overline{\bf J}$.
\label{fig3}}\end{minipage}
\hspace{1pc}%
\begin{minipage}{17.5pc}
\includegraphics[height=2.25in]{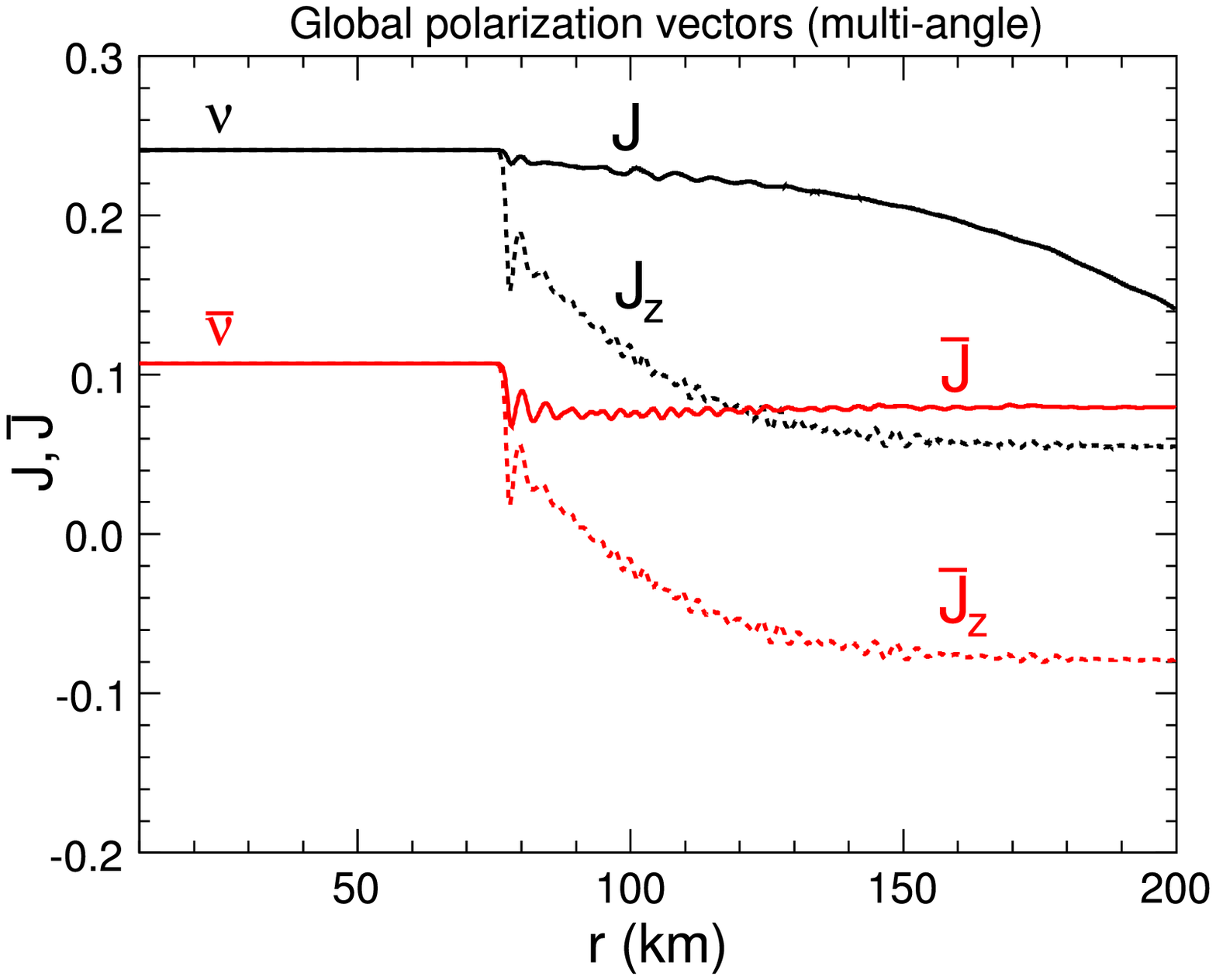}
\caption{ Multi-angle simulation in inverted hierarchy: modulus and $z$-component of $\bf J$ 
and $\overline{\bf J}$.
\label{fig6}}
\end{minipage} 
\end{figure}
 
Figures~\ref{fig3} and \ref{fig6} show the evolution of the
global polarization vectors (modulus $J$ and
$z$-component $J_z$) for neutrinos and antineutrinos, in the single- and multi-angle cases.
The behavior of these vectors can be related to the gyroscopic pendulum motion. 
At the beginning, in the synchronized regime, all the
polarization vectors are aligned so that  $J=J_z$ and $\overline J=\overline J_z$: the
pendulum just spins in the upward position without falling. Around $\sim 70$ km
the pendulum falls for the first time and nutations appear. The nutation amplitude gradually decreases 
and bipolar oscillations eventually vanish for $r\sim 100$~km. At the same time, the spectral split builds up:
antineutrinos tend to completely reverse their polarization, while this happens only partially for
neutrinos. As said before, also for antineutrinos there is a partial swap of the spectra for $E\sim 4$~MeV.
From Figure~\ref{fig6}  it appears that 
bipolar oscillations of $\bf J$ and $\overline{\bf J}$ are largely smeared out in the multi-angle case.
The bipolar regime starts somewhat later with respect to the single-angle
case,  since neutrino-neutrino interaction angles can be larger than the (single-angle) average one,
leading to stronger self-interaction effects, 
that force the system in synchronized mode  slightly longer.
However, just as in the single-angle case, the spectral split builds up, $\overline{J}_z$ gets finally reversed, while
the difference $D_z=J_z-\overline{J}_z$ remains constant.

\begin{figure}[t]
\center
\includegraphics[height=2.4in]{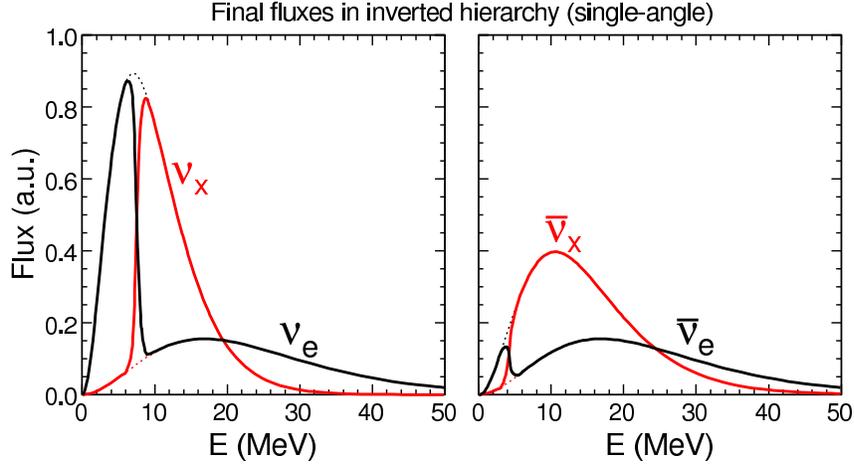}
\caption{Single-angle simulation in inverted hierarchy: 
final fluxes (at $r=200$~km, in arbitrary units) 
for different neutrino species as a function of energy. Initial fluxes are shown
as dotted lines.
\label{fig5}}
\end{figure}

\begin{figure}[h]
\center
\includegraphics[height=2.4in]{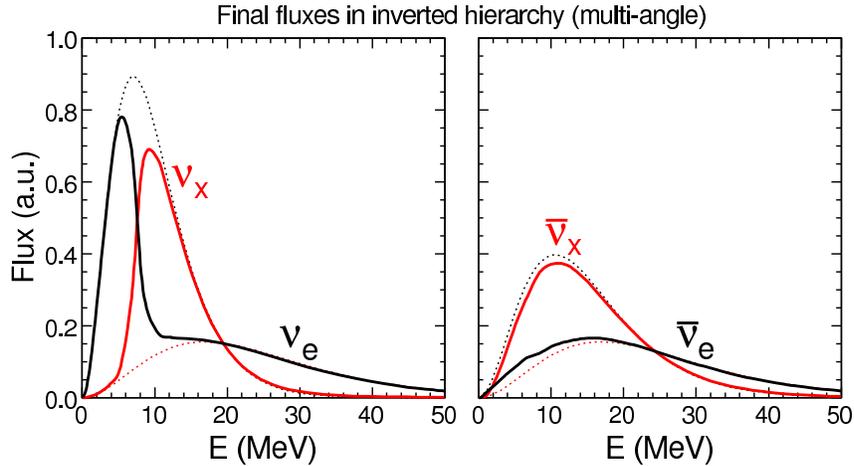}
\caption{As in Fig.~7, but for multi-angle simulations.
\label{fig8}}
\end{figure}

Figures~\ref{fig5} and \ref{fig8} show the final $\nu$ and 
$\overline\nu$ fluxes, in 
the single- and multi-angle simulations.
The $\nu$ clearly show the spectral split effect and the 
corresponding sudden swap of $\nu_e$ and $\nu_x$ fluxes above $E_c\simeq 7$~MeV.
In the right panel of Figure~\ref{fig5}, the final 
$\overline\nu$ spectra are basically
completely swapped with respect to the initial ones, except at very low energies, where there 
appears a minor $\overline\nu$ spectral split.
This phenomenon can be related to the loss of $\overline J$ 
and of $|\overline{J}_z|$~\cite{ourCollective}.
Also in the multi-angle case of Figure~\ref{fig8} ,
the $\nu$ spectral swap at $E>E_c\simeq7$~MeV is rather
evident, although less sharp  with respect to
the single-angle case, while the
minor feature associated to the ``antineutrino spectral split''  is largely smeared out. We conclude that at least the $\nu$ spectral split (left panel
of either Fig.~7 or 8) provides a robust and potentially observable
collective feature emerging in numerical experiments for inverted hierarchy.
\vspace*{0mm}
\newpage

\section{Summary}

We have studied supernova neutrino oscillations in a model where the collective flavor transitions
(synchronization, bipolar oscillations, and spectral split) are
well separated from later, ordinary MSW effects.
We have performed numerical simulations in both single- and multi-angle
cases, using continuous energy spectra with significant $\nu$-$\overline \nu$
and $\nu_e$-$\nu_x$ asymmetry. The results
of the single-angle simulation can be largely understood
by means of an analogy with a classical gyroscopic pendulum. 
The main  observable effect appears to be 
the swap of final-state energy spectra, for inverted hierarchy,
at a critical energy dictated by lepton number conservation.
In the multi-angle simulation, details of
self-interaction effects can change,
but the spectral split remains a robust, observable feature. In this sense,
averaging over neutrino trajectories does not alter the main effect of the self interactions.
From the point of view of neutrino parameters, collective flavor oscillations in 
supernovae could be instrumental in identifying the inverse neutrino mass hierarchy, even for tiny 
$\theta_{13}$.~\cite{DigheMiri}

\begin{figure}[b]
\center
\includegraphics[height=1.9in]{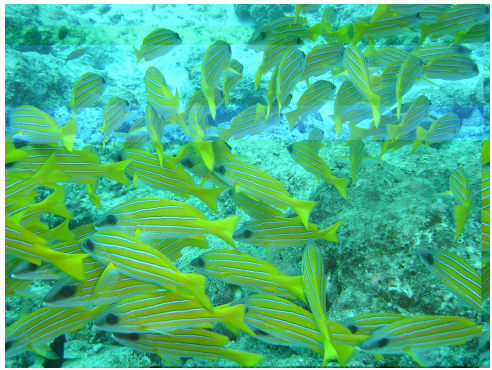}
\caption{A school of fish branching out in two different directions. 
Analogously, supernova neutrino polarization vectors might split up in flavor space, due to self-interaction effects.} 
\label{school2}
\end{figure}

\section{Epilogue}

Supernova $\nu$ and $\overline\nu$ flavor polarization vectors can perform elaborate
and collective ``dances'' (precession, nutations) in flavor space,
at least for the first $O(100)$~km, in the case
of inverted mass hierarchy. Many aspects of this behavior, however,
seem to be precluded to experimental observations, with the possible exception
of a robust, finale-state feature: the $\nu$ spectral split. 
According to a calculable energy threshold, supernova $\nu$ might
then proceed to the Earth with their original flavor (say, $\nu_e$), 
or with the complementary one (say, $\nu_x$), just as an initially
coherent school of fish (Fig.~\ref{school}) may finally 
branch out in some circumstances (Fig.~\ref{school2}).

\newpage
\section{Acknowledgements}
E.L.\ thanks Milla Baldo Ceolin for kind hospitality at
the NO-VE 2008 Workshop in  Venice, where these results were presented. 
This work is supported in part by the Italian INFN and MIUR through the ``Astroparticle Physics'' research project.

\end{document}